\documentclass[aps,prl,showpacs,twocolumn,groupedaddress,amsmath,amssymb]{revtex4}
\usepackage{times,xspace}
\usepackage{amsfonts}
\usepackage{amsfonts}
\usepackage{bm}
\usepackage{amssymb}
\usepackage[final]{graphicx}
\usepackage{color}
\begin{document}

\title{Two energy scales in the magnetic resonance spectrum of electron and hole doped pnictide superconductors}

\author{Tanmoy Das$^1$, and A. V. Balatsky$^{1,2}$}
\affiliation{$^1$Theoretical Division, Los Alamos National Laboratory, Los Alamos, NM-87545, USA.\\
$^2$Center for Integrated Nanotechnology, Los Alamos National Laboratory, Los Alamos, NM-87545, USA.}

\date{\today}
\begin{abstract}
We argue that a multiband superconductor with sign-changing gaps may have multiple spin resonances. We calculate the RPA-based spin resonance spectra of a pnictide superconductor by using the five band tight-binding model or angle-resolved photoemission spectroscopy (ARPES) Fermi surface (FS) and experimental values of superconducting (SC) gaps. The resonance spectra split in both energy and momenta due to the effects of multiband and multiple gaps in $s^{\pm}-$pairing; the higher energy peak appears around the commensurate momenta due to scattering between $\alpha-$FS to $\gamma/\delta-$FS pockets.  The second resonance is incommensurate coming from $\beta-$FS to $\gamma/\delta-$FS scatterings and its $q-$vector is doping-dependent and hence on the FS topology. Energies of both resonances $\omega^{1,2}_{res}$ are strongly doping dependent and are proportional to the gap amplitudes at the contributing FSs. We also discuss the evolution of the spin excitation spectra with various other possible gap symmetries, which may be relevant when either both the electron pockets or both the hole pockets completely disappear.
\end{abstract}
 \maketitle\narrowtext
{\it Introduction:} The magnetic resonance behavior which is directly probed by the inelastic neutron scattering (INS) spectroscopy
gives valuable information about the pairing mechanism of the unconventional superconductors. In cuprates, INS exhibits a clear signature of a resonance mode in addition to its characteristic dispersion (known as `hour-glass' behavior) which is enhanced dramatically below $T_c$, and the mode energy scales universally with the SC gap amplitude.\cite{greven} In a multiband unconventional superconductor, the situation becomes more complex due to the presence of the multiband and multiple SC gaps as well as the possibility of having multiple pairing symmetries. Given the important role the spin resonance played in the identification of pairing symmetry in the past, we investigate the details of spin resonance in a multiband model relevant for pnictide superconductors. The primary question we ask: {\em Could  multiple superconducting gaps at various bands lead to multiple spin resonances} as opposed to the case of single spin resonance for single band superconductors? We find that indeed multiband superconductors could have multiple spin resonances. It is also important to study how the resonance spectra evolve in $({\bf q},\omega)-$phase space as function of doping and other parameters.

An example of a multiband superconductor is MgB$_2$. It is a conventional $s-$wave superconductor and its signal in the INS measurement is absent as the magnetic spectra are sensitive to the sign change of the SC gap on the FS sheet.\cite{chen} Recently discovered pnictides offer a new testing ground for these ideas.\cite{kamihara,gang} In pnictides, it is now generally accepted that an unconventional $s^{\pm}$ pairing symmetry is present with evidence coming  from ARPES\cite{nakayama} and other probes\cite{christiansonreview,johnston}. This conclusion is also supported by theoretical calculations\cite{maier,chubukov}.

We have performed a standard BCS susceptibility calculation within the RPA framework with $s^{\pm}-$wave pairing symmetry for the materials with the Fe-122 band structure. The computed dispersion and intensity of the spin polarized INS spectra reveal the following: (i) Two SC gaps set the scale for the two spin resonances $\omega^1_{res}$ and $\omega^2_{res}$ that arise from the fact that the SC gap magnitudes (although the same pairing symmetries) are different at two hole pockets at $\Gamma$. At the same time the momenta for spin resonances are different: $\omega_{res}^1$ is commensurate and $\omega^2_{res}$ is incommensurate (with doping-dependent $q-$vectors). (ii) We also find that both $\omega^1_{res}$ and $\omega^2_{res}$ are also doping dependent but universally proportional to the SC gap amplitudes. When one of the hole pockets disappears, one of the resonances also disappears. (iii) Furthermore, we also show that when both hole pockets or both electron pockets vanish completely from the FS, the pairing symmetry should be changed to one of the form which will give a change of sign at the same FS and as a result the intraband scattering spectra will be shifted to the low-energy and low-momenta scale.\cite{chubukov}

The large energy scale of both modes $\omega^1_{res} \sim 8- 15$meV and $\omega^2_{res} \sim 5-12$meV depending on doping makes it likely that {\em both} resonances can be seen in INS. There is clear evidence for the first commensurate resonance\cite{christianson,Dai} and there is also some evidence for the second mode\cite{lumsden,christianson,Dai} whose energy and ${\bf q}-$values agree well with our calculation.

{\it Formalism:} The bare dispersion and its corresponding orbital characters are calculated by solving the five $d-$orbitals (in one Fe unit cell) tight-binding formalism.\cite{graser}
We obtain doping by the rigid band shift method.
In a multiorbital SC state, the BCS free-fermion susceptibility is a tensor\cite{maier} given by
\begin{eqnarray}\label{eq:1}
&&\chi_{0rstu}({\bf q},i\omega_m)=-\frac{1}{2}\sum_{k,n,\nu,\nu^{\prime}}M^{\nu\nu^{\prime}}_{rstu}({\bf k},{\bf q})[G^{\nu}_{\bf k}(i\omega_n)\nonumber\\
&&\times G^{\nu^{\prime}}_{{\bf k}+{\bf q}}(i\omega_n+i\omega_m)+F_{\bf k}^{\nu}(i\omega_n)F_{-{\bf k}-{\bf q}}^{\nu^{\prime}}(-i\omega_n-\omega_m)].
\end{eqnarray}
Here $\nu$ ($\nu^{\prime}$) is the initial (final) eigenstate coming from $s,t$ ($r,u$) orbitals and $G$ and $F$ are the normal and anomalous part of the Green's function, respectively. $M$ is the orbital to band matrix element (ME) made of the eigenvectors. The many-body poles in the dressed $\chi$ are obtained within the RPA framework. The terms in the spin RPA interaction vertex $U_s$ that are included in the present calculation are intraorbital interaction $U$, an interorbital interaction $U^{\prime}$, Hund's rule coupling $J>0$, and the pair hopping strength $J^{\prime}$. The rest of the interaction can be expressed in term of $U$ as $U^{\prime}=U-J/2$, $J=U/4$ and $J^{\prime}= J$.\cite{maier} While the calculations are done in the unfolded one Fe per unit cell Brillouin zone (BZ), the results are subsequently folded to the two Fe per unit cell BZ.

\begin{figure}[h]
\hspace{-0cm}
\rotatebox{0}{\scalebox{.47}{\includegraphics{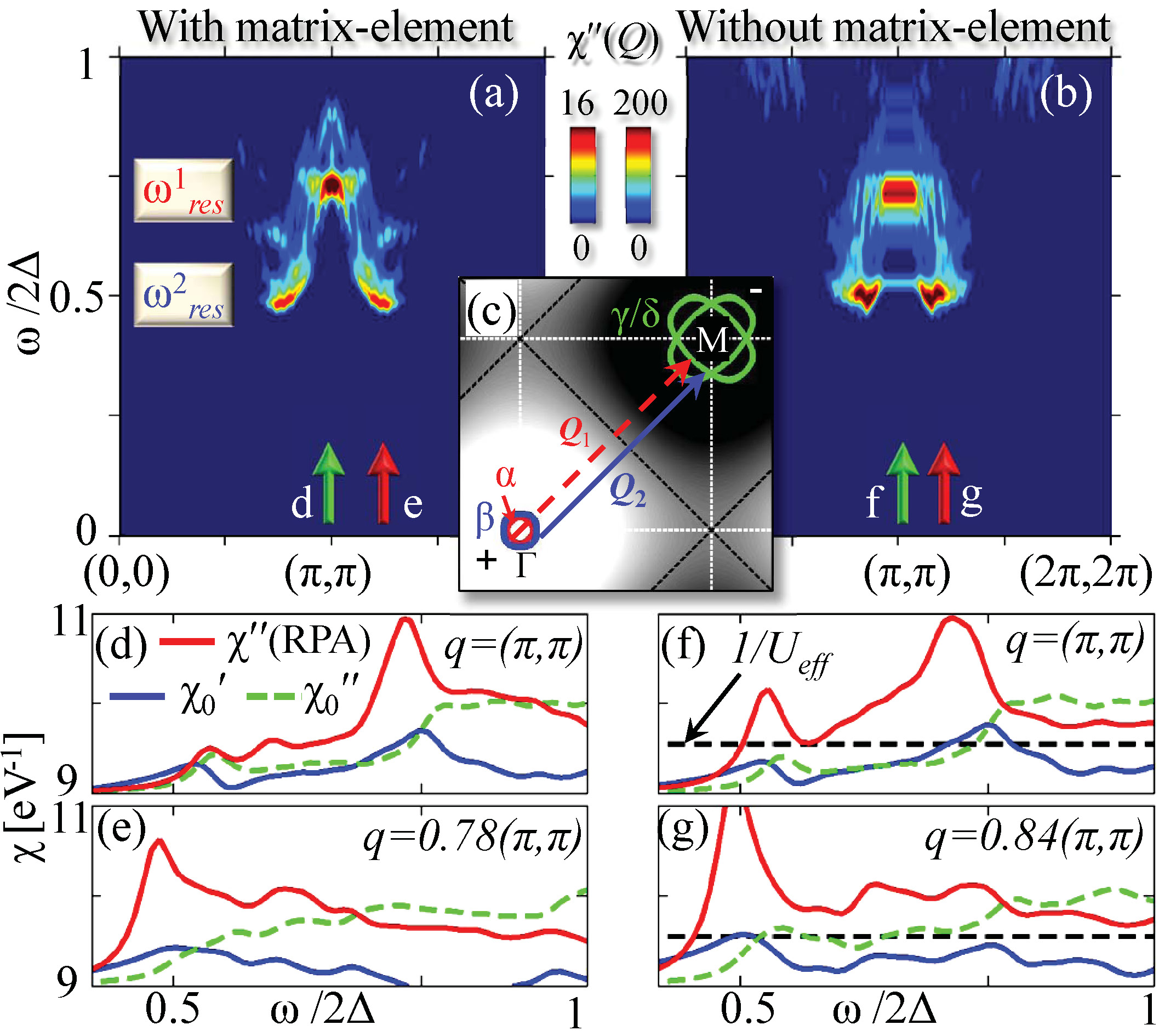}}}
\caption{(Color online)
(a)-(b) RPA-BCS $\chi^{\prime\prime}(\omega)$, plotted for calculation including the ME effect, $M$ from Eq.~1 in (a) and without including the ME effect ($M=1$) in (b). (d)-(g) The corresponding lower panel figures clarify the mechanism of double resonances at two representative momenta cuts. $\chi^{\prime\prime}$s are shifted by a constant value to ease comparison with others, while they all go to zero at $\omega=0$. In (f)-(g), when $1/U_{eff}$ intersects with $\chi_0^{\prime}$, a true resonance occurs within the RPA. In (d)-(e), the $1/U$ in the $U_s$ vertex does not directly intersect to total $\chi_0^{\prime}$, but to its intraorbital components only and thus not shown.\cite{graser} (c) The corresponding LDA FS in the folded zone is plotted at the same doping. The white to black background color gives the $s^{\pm}-$symmetry. The two arrows point to the nesting channels which are responsible for the two modes in (a). The calculation is done with SC gap $\Delta_{\alpha}=\Delta_{\gamma/\delta}=2\Delta_{\beta}=7$meV.} \label{fig1}
\end{figure}
{\it Double resonances:} Fig. 1 shows our computed magnetic spectrum as a function of energy along the diagonal direction in the SC state for an representative doping of $x=0.20$. The essential ingredient for the appearance of two resonances is having different amplitudes of gaps on different pieces of FSs. ARPES measures the average SC gaps which follows roughly for all dopings as $\Delta_{\alpha}\approx\Delta_{\gamma/\delta}\approx2\Delta_{\beta}$, where $\alpha$ and $\beta$ are the inner and outer hole pockets at $\Gamma$, respectively and $\gamma/\delta$ are the electron pockets at $M=(\pi,\pi)$ as shown in Fig.~1(c).

The superconductivity is included in BCS formalism with $s^{\pm}-$pairing symmetry as $\Delta_{i}({\bf k})=\Delta_i[\cos{(k_xa)}+\cos{(k_ya)}]/2$, where $i$ is the band index. With these experimental inputs, we find a clear signature of two resonance modes: $\omega^{1}_{res}\sim0.75\times(2\Delta_{\alpha})$ near the commensurate momentum $Q=(\pi,\pi)$ and the second one at $\omega^{2}_{res}\sim0.5\times(2\Delta_{\alpha})<\omega^{1}_{res}$ at an incommensurate vector $q\sim0.78(\pi,\pi)$. This is the main result of this paper.

According to the conventional view, apart from the ME effect the magnetic structure in BCS $\chi_0^{\prime}$ is entirely governed by the sign change of gaps at the `hot-spots' and the energy conservation formula for inelastic scattering of Bogoliubov quasiparticle on the FS\cite{norman}
\begin{eqnarray}\label{eq:3}
\omega^{\nu\nu^{\prime}}({\bf k}_F, {\bf q})=|\Delta^{\nu}_{{\bf k}_F}|+|\Delta^{\nu^\prime}_{{\bf k}_F+{\bf q}}|.
\end{eqnarray}
At this locus on the $({\bf q},\omega)-$phase space, $\chi_0^{\prime}$ attains a logarithmic jump, and due to the Kramers' Kronig relation $\chi_0^{\prime\prime}$ have a discontinuous peak as shown by the blue solid and green dashed lines in Figs.~1(d-e). Therefore within the RPA, for a broad range of interaction $U$, a resonance is possible at the same locus.\cite{chubukov} With these observations, apart from the true intensity of $\chi_0^{\prime}$, all its dispersion in the magnetic spectrum can be understood from the spanning vectors ${\bf q}$s between various FS pieces which have opposite signs of the SC gap.

At the doping considered in Fig.~1, all four pockets are present on the FS; see Fig.~1(c). For the case of $s^{\pm}-$pairing, the gap below the magnetic BZ is always positive and above that, it is negative; i.e., $\alpha$ and $\beta$ FS always have a positive sign while $\gamma$ and $\delta$ bands have a negative sign at all dopings studied here, as illustrated by the white to black color background in Fig.~1(c). Thus only the inelastic scattering between $\alpha,\beta\leftrightarrow\gamma,\delta$ is allowed, and the others including intraband scattering are prohibited. The $\alpha-$FS piece resides closer to the $\Gamma-$point than the $\beta$ one, and $\gamma/\delta$ lies close to $M=(\pi,\pi)$ points. Therefore the corresponding magnetic vector $Q_{1}$ for the scattering channel $\alpha\leftrightarrow\gamma,\delta$ is larger than $Q_2$ for $\beta\leftrightarrow\gamma,\delta$.

To clarify the physical origin of the two energy scales within the RPA, we take two representative ${\bf q}$ cuts for $\chi^{\prime}_0$, $\chi^{\prime\prime}_0$ and $\chi^{\prime\prime} (RPA)$, Fig.~1(d) and (e). At $Q=(\pi,\pi)$ in Fig.~1(d), $\chi^{\prime}_0$ shows two characteristic peaks at $\omega^{1}_{res}$ and $\omega^{2}_{res}$ coming from $Q_1$ and the residual $Q_2$ scatterings respectively. The former peak is stronger than the latter as the corresponding vector $Q_1$ spans up to $Q$ but $Q_2<(\pi,\pi)$. The situation is reversed at $q=0.78(\pi,\pi)$ where  $Q_1>0.78(\pi,\pi)$ but $Q_2\sim0.78(\pi,\pi)$, and thus the lower energy peak gains more intensity while the higher energy one goes negative, prohibiting the possibility of any resonance to occur within the RPA.

In the RPA, the resonance peaks are set by the locus [Eq.~2] and intensity (ME in Eq~1) of the discontinuities of $\chi^{\prime}_0$. We use $U=1.1$eV, a choice that is reasonable; moreover, the results are representative and remain qualitatively the same for different $U$. The value of $U$ is close to the value of 1.2eV used in an earlier calculation to produce one resonance, which is closer to the first resonance we find here.\cite{maier} With this choice of $U$, the higher energy peak in $\chi^{\prime}_0$ yields resonance peak at $[Q,\omega^{1}_{res}]$, which is clearly separated from the second resonance at $[0.78(\pi,\pi),\omega_{res}^2]$. Away from these two sharp resonance peaks, the intensity is spread out over the large $({\bf q},\omega)$ region, showing a remarkable downward dispersion as shown Fig.~1(a). The downward dispersion in pnictide can be contrasted to the upward dispersion seen in chalcogenides.\cite{argyriouupward}

\begin{figure}[top]
\hspace{-0.5cm}
\rotatebox{0}{\scalebox{.4}{\includegraphics{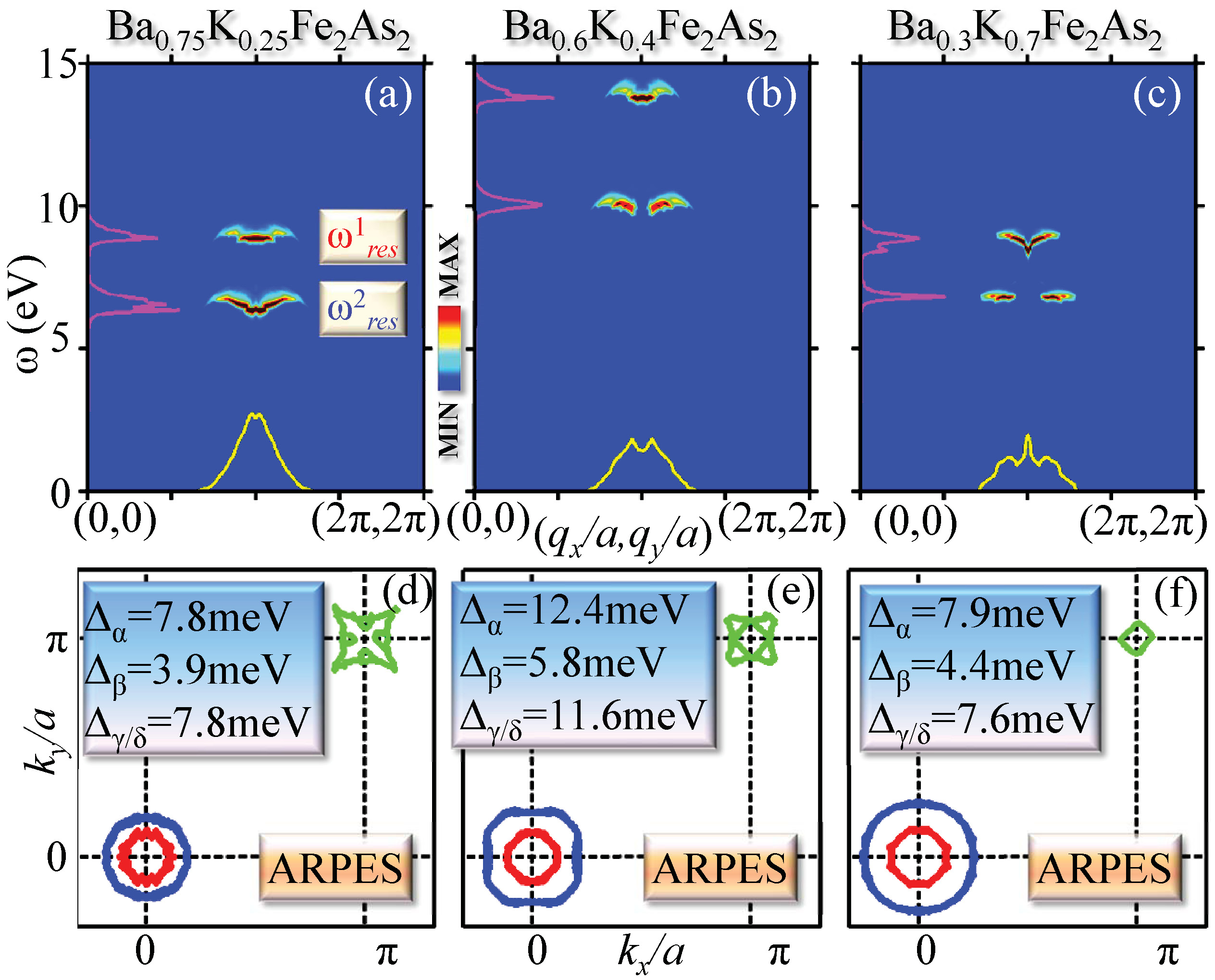}}}
\caption{(Color online)
(a)-(c) Computed spectra of $\chi^{\prime\prime}({\bf q},\omega)$ are plotted in color for underdoping in (a), optimal doping in (b) and overdoping  in (c) of the hole-doped sample.  The calculation is done here within in the minimal model described in the text. The magenta and yellow lines are the ${\bf q}$ and $\omega$ integrated value of $\chi^{\prime\prime}$, respectively. The two energy scales in (b) match well with the available data at the same doping in Ref.~\onlinecite{christianson}. (d)-(f) The ARPES FSs corresponding to the INS spectra given in their upper panel at the same dopings\cite{nakayama}.
} \label{fig1}
\end{figure}

In the rest of the paper, we introduce a minimal model calculation of the INS spectra using ARPES or LDA FS which may still be valid in a multiband system even when the ME effects are neglected [$M=1$ in Eq.~1]. We investigate the consequences of avoiding ME effect in Figs.~1(b) and 1(f-g). $\chi^{\prime}_0$ shows qualitative similarity between the calculation with and without including ME [compare Fig.~1(d-e) with Fig.~1(f-g)]: While both the peaks are still present, the locations of their maxima are shifted slightly in both $q$ and $\omega$ space. But the relative intensity of the peaks is subject to change, which can be compensated by tuning the effecting interaction $U_{eff}$. Technically, setting the ME to be 1, we get all components of $\chi$ tensor to be equal in Eq.~1. Approximating the RPA interaction vertex $U_s$ by an effective scalar interaction $U_{eff}<U$, we can simplify the total RPA $\chi=\sum {\tilde{\chi_0}(\tilde{1}-\tilde{U_s}\tilde{\chi_0})^{-1}}\approx\chi_0(1-U_{eff}\chi_0)^{-1}$ (the symbol tilde over a quantity signifies that it is a tensor). We find that $U_{eff}=0.1$eV can reasonably reproduce the results obtained with the ME effect including both resonances and the downward dispersion of the spectra. Of course this approximation does not preclude us from analyzing the full problem but merely simplifies the analysis.
\begin{figure}[top]
\hspace{-0.5cm}
\rotatebox{0}{\scalebox{.4}{\includegraphics{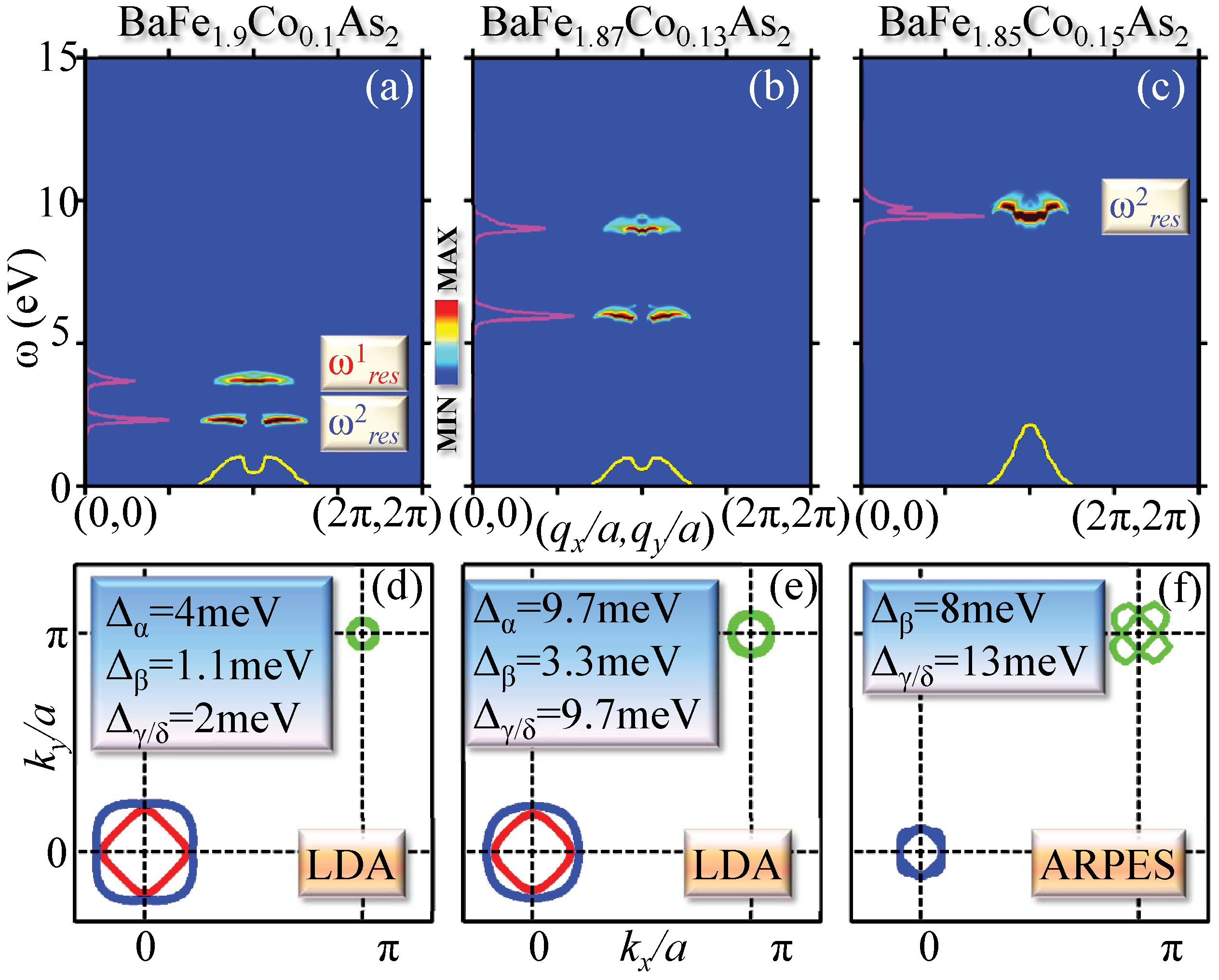}}}
\caption{(Color online) Computed INS spectra plotted for two underdopings in (a,b) and optimal doping in (c) of the electron doped case. (d)-(f) The corresponding FSs are given at their lower panels at the same dopings obtained from the LDA band in (d) and (e) and from ARPES data\cite{terashima} in (f). The gap values in (d) and (e) are available from penetration depth measurement\cite{luan}, and optics\cite{perucchi} respectively at the same dopings where the same values for (c) are obtained from ARPES\cite{terashima}. In the LDA, FS pockets are known to be slightly larger than the experimental value due to the neglect of the spin density wave order.\cite{marty}.} \label{fig1}
\end{figure}

In this approach, we compute the INS spectra with the RPA correction using the FS obtained either from ARPES whenever available or from the LDA. The other input in our calculation is the gap values which we obtain from various probes as listed in the corresponding figures below. An added benefit of this approach is that it allows us to compute the INS spectrum entirely from experimental inputs, which is one of our goals.

{\it Doping dependence of two resonances:} We compute the doping evolution of the two resonances by using experimental SC gaps and the ARPES FS. We note that while the doping dependence of the two resonances can also be obtained from the full BCS-RPA calculation (not shown), the energy and ${\bf q}$ values of the resonances are determined more accurately from the ARPES FS due to its inconsistency with LDA FS.

Fig.~2 shows our computed spin resonance spectrum along the diagonal direction in the SC state as a function of dopings in hole-doped Fe-122 compounds. The separation between the two resonances in both energy and momenta is present for the doping range when both the hole pockets at $\Gamma$ are present.


The first mode, $\omega_{res}^1$, occurs at $Q=(\pi,\pi)$ at all dopings with a dispersion branch which is characteristic of the shape of the $\alpha-$FSs. On the contrary, the doping dependence of the second mode $Q_2$ is significantly large, deviating from a commensurate to incommensurate spectra as the contributing $\beta-$FS grows in size with hole doping. We predict that with further underdoping, the second mode persists even when the first mode may disappear at a doping when the $\alpha-$FS vanishes (see electron doping below). Furthermore, we predict that with overdoping, when the electron pockets at $M$ disappear, both these resonances will vanish. And, if the pairing is still present, the corresponding gap symmetry will change to the one for which the gap will change sign in the same FS pieces (such as $d_{x^2-y^2}$ or $d_{xy}$); the neutron mode will be confined to the small ${\bf q}$-space (see supplementary material).

For electron doping the ${\bf q}$ dependence of the second mode is opposite to the hole doping as the doping evolution of its FS topology is opposite to that of the hole-doped one, see Fig.~3. With increasing electron doping, the hole pockets at $\Gamma$ gradually disappear while the electron pockets grow in size. The spatial separation between the two modes gradually decreases as the distance between the $\alpha-$ and $\beta-$FSs decreases and both modes tend to be more commensurate in nature than that in the higher doping of the hole-doped side. This behavior is consistent with the reasons why the static commensurate spin-density wave order is more stable in electron-doping than in the hole doped one.\cite{marty,christiansonreview,johnston} Finally, when the $\alpha$ band vanishes from the FS, $\omega_{res}^1$ also vanishes; see Fig.~3(c).

For electron doping, even when the first resonance vanishes at $x=0.15$, the second resonance is still present at $\omega_{res}=9.3$meV at $Q$ for the ARPES FS and the experimental gap amplitudes and agrees reasonably with the experimental value of $8.6$meV at $Q$ at a slightly larger doping of $x=0.16$ [Ref.~\onlinecite{lumsden}]. And with further electron doping, the $\beta-$FS also vanishes and the second resonance disappears. In this doping the superconductivity is seen to disappear.\cite{sekiba}

\begin{figure}[h]
\hskip-0cm
\rotatebox{0}{\scalebox{.4}{\includegraphics{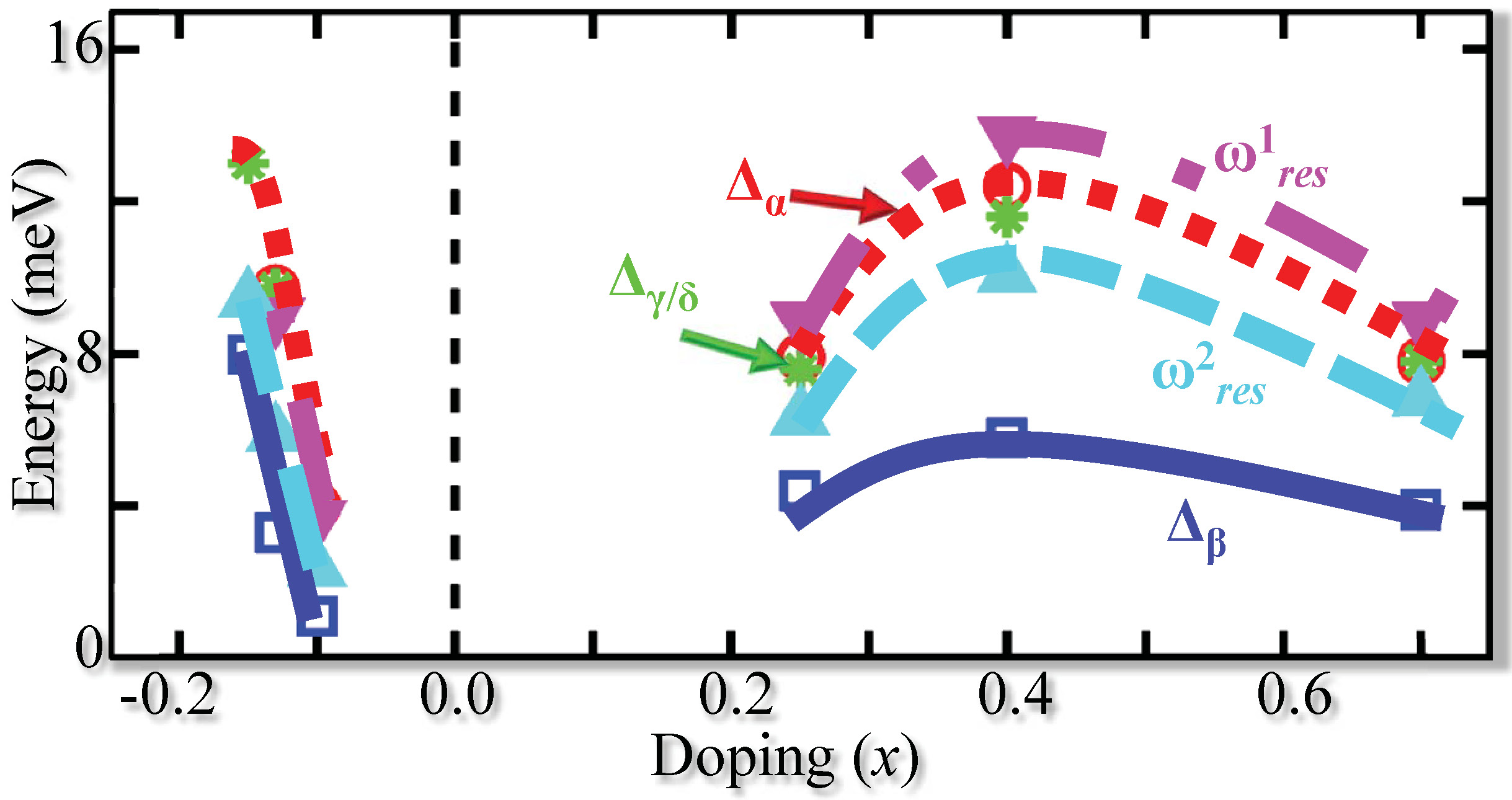}}}
\caption{{(Color online) Doping dependence of two resonance energies and the SC gap amplitudes.} The open symbols of various colors gives the SC
gap values [red circles=$\Delta_{\alpha}$, blue squares=$\Delta_{\beta}$, green stars=$\Delta_{\gamma/\delta}$] as taken from experiments (see text).
The filled symbols are the two resonance energies [magenta lower (cyan upper) triangles=$\omega_{res}^1$ ($\omega_{res}^2$)], shown in Figs.~1,2. The shadings of same colors are a guide to the eyes.}  \label{fig3}
\end{figure}

The two modes are not only ${\bf q}-$resolved, they also appear at two energy scales as their underlying scattering states have different gap values. According to Eq.~2, each of the INS spectra is directly related to the SC gap values of participating scattering states. For example, the resonance at $Q_1$,  $\omega^{1}_{res}=\Delta_{\alpha}|g^{\alpha}_{{\bf k}_F}|+\Delta_{\gamma/\delta}|g^{\gamma/\delta}_{{\bf k}_F+Q_1}|\approx2\Delta_{\alpha}g^{\alpha}_{{\bf k}_F}$, by using the fact that $\Delta_{\alpha}\approx \Delta_{\gamma/\delta}$ and $g^{\alpha}_{{\bf k}_F}=-g^{\gamma/\delta}_{{\bf k}_F+{\bf q}}$. Therefore, in pnictide the leading or the higher energy resonance is always $\omega^1_{res}\propto 2\Delta_{\alpha}$ at all dopings (see Fig.~4). The proportionality  constant is doping dependent, set by the value of the `hot-spot' momenta. At optimal doping $x=0.4$ of the hole doped case, we find that $\omega_{res}^1=13.7$meV and $\omega_{res}^2=10$meV by using the ARPES FS and gap values which agree remarkably well with the experimental value of 14 and 9.5meV, respectively at the same doping\cite{christianson}. There is no one-to-one correspondence of the resonance energy with $T_c$ as $2\Delta/k_BTc$ is found to be doping dependent\cite{nakayama}.

This universal relationship between the resonance energy and SC gaps indicates that pnictides are in the weak- or intermediate-superconducting-coupling region where mean-field treatment of the SC state works well. Interestingly, the intensity of the resonance is found experimentally to follow the same $T$ dependence of the SC gaps and vanishes at $T_c$ in pnictide\cite{Dai} and chalcogenides\cite{mook}. The universal relationship $\omega_{res}\propto(\Delta_1+\Delta_2)$ (where $\Delta_1$, and $\Delta_2$ are the SC gap amplitudes of the two orbitals connected by the `hot-spot') is also found to be consistent with other families of superconductors.\cite{greven}

{\it Conclusion:} In conclusion, we have shown that for a multiband system with unconventional pairing with sign-changing gaps, one expects to find multiple spin resonances which are separated both in energy and momenta. The essential mechanism for having two resonances in pnictide is the splitting between two hole pockets with different gap amplitudes. The difference in SC gaps and $q-$vectors of these two bands are the main reason why we see two resonances. Any calculation invoking two bands will thus miss the second resonance. These results can be compared with INS experiments with the aim to observe second resonance in pnictides.

The present computation does not have any free parameter as both the required FS information is obtained from either ARPES or the LDA and the SC gap values are taken from experiments. Our method of analysis reveals an intriguing relationship between the ARPES and INS measurements and one spectroscopic information can be extracted from the other one.  This will give an important tool for any other multiband superconductors to study the pairing symmetry and/or the FS information which are still not clear.

\begin{acknowledgments}
We are grateful to A. Christianson, R. S. Markiewicz and A. Bansil for useful discussions. This work is funded by US DOE, BES and LDRD.
\end{acknowledgments}


\begin{thebibliography}{99}
\bibitem{greven} G. Yu, {\it et al.}
Nat. Phys. {\bf 5}, 873 (2009).
\bibitem{chen} X. K. Chen, {\it et al.}
%
%
Phys. Rev. Lett. {\bf 87}, 157002 (2001).
%
\bibitem{kamihara}Y. Kamihara, {\it et al.}
J. Am. Chem. Soc. {\bf 130}, 3296 (2008).
%
\bibitem{gang}Mu Gang, {\it et al.}
Chinese Phys. Lett. {\bf 25}, 2221 (2008).
%
%
\bibitem{nakayama}K. Nakayama, {\it et al.}
arXiv:1009.4236.
%
\bibitem{christiansonreview} M. D. Lumsden,  {\it et al.} J. Phys.: Cond. Mat. {\bf 22}, 203203 (2010).
%
\bibitem{johnston} D. C. Johnston, arXiv:1005.4392.
%
\bibitem{maier}T. A. Maier, {\it et al.}
Phys. Rev. B {\bf 79}, 134520 (2009).
%
\bibitem{chubukov}A. V. Chubukov {\it et al.} Phys. Rev. B {\bf 78}, 134512 (2008);
 M. M. Korshunov, and I. Eremin, Phys. Rev. B {\bf 78}, 140509(R) (2008).
%
\bibitem{christianson} A. D. Christianson, {\it et al.}
%
Nature {\bf 456}, 930 (2008).
%
\bibitem{Dai}C. Zhang, {\it et al.}
%
arXiv:1012.4065.
%
\bibitem{lumsden}M. D. Lumsden, {\it et al.}
%
Phys. Rev. Lett. {\bf 102}, 107005 (2009).
%
%
\bibitem{graser}S. Graser, {\it et al.}
Phys. Rev. B {\bf 81}, 214503 (2010).
%
\bibitem{norman} I. Eremin {\it et al.}
Phys. Rev. Lett. {\bf 94}, 147001 (2005).
%
\bibitem{argyriouupward}D. N. Argyriou {\it et al.},
Phys. Rev. B {\bf 81}, 220503(R) (2010).
\bibitem{terashima}K. Terashima,{\it et al.}
Proc. Nat. Acad. Sci. USA {\bf 106}, 7330 (2009).
%
%
\bibitem{luan} L. Luan, {\it et al.}
Phys. Rev. B {\bf 81}, 100501(R) (2010).
%
\bibitem{perucchi}A. Perucchi, {\it et al.}
Eur. Phys. J. B {\bf 77}, 25 (2010).
%
\bibitem{marty} K. Marty, {\it et al.}
%
arXiv:1009.1818.
%
\bibitem{sekiba}Y. Sekiba, {\it et al.}
%
New J. Phys. {\bf 11}, 025020  (2009).
%
\bibitem{mook}H. A. Mook, {\it et al.}
arXiv:0904.2178.
%
\end{thebibliography}
\end{document}